\documentclass[aps, twocolumn, showpacs,superscriptaddress]{revtex4-1}
\usepackage[english]{babel}

\usepackage{graphics}
\usepackage{amsmath}
\usepackage{graphicx}
\usepackage[colorlinks=true, allcolors=blue]{hyperref}
\usepackage{lipsum}

\begin{document}
	
	\title{Effective cross-plane thermal conductivity of  metal-dielectric multilayers at low temperatures}

	\author{A.~I.~Bezuglyj}
	
	\affiliation{\it National Science Center "Kharkiv Institute of Physics and Technology",   \\
		1, Akademichna St., Kharkiv, 61108, Ukraine\\}

	\author{I.~V.~Mironenko}
	\affiliation{\it Kharkiv National University, 4, Svobody Sq., Kharkiv, 61022, Ukraine }
	
	\email[Corresponding author: ]{mironenko@karazin.ua}
	
	\author{V.~A.~Shklovskij}
	\affiliation{\it Kharkiv National University, 4, Svobody Sq., Kharkiv, 61022, Ukraine }


	\begin{abstract} 
		
		Heat transfer in layered metal-dielectric structures is considered theoretically based on an analytical solution of the Boltzmann transfer equation for the phonon distribution function. Taking into account the size effect, the problem of effective cross-plane thermal conductivity of structures containing two metal layers is analyzed in detail. If the thickness of the metal layers is less than the phonon mean free path, interlayer heat transfer is carried out predominantly by phonons, and the effective cross-plane thermal conductivity is determined by the reflection of phonons from the metal/dielectric interfaces. In the opposite case of thick metal layers, the effective cross-plane thermal conductivity is determined both by the thermal conductivity of the metal layers and by the thermal resistance of the dielectric layers. The results obtained are generalized to multilayer structures and superlattices. 
	\end{abstract}
	
	\maketitle

	\section{Introduction}\label{s1}
	
	The properties of layered structures and the control of heat fluxes in such structures have been areas of active research in recent decades.
	The miniaturization of the components of modern electronic devices and the increase in their operating frequency has led to an increase in the density and intensity of heat sources and, consequently, to the problem of efficient heat removal.\cite{Pop2010, Cahill_review2014, Giri2020} In multilayer systems containing layers of a ferromagnet, heat transfer plays an important role in the spin Seebeck effect and in the entire field of spin caloritronics.\cite{Bauer2012, Schreier2013b, Boona2014} Layered metal-semiconductor structures are the basis for plasmonic devices and heating can have a positive effect on the performance of such devices.\cite{Saha2018, Mascaretti}
	The problem of reducing the thermal conductivity of layered structures is of interest since it is important for applications that require minimization of the heat flow between layers (thermoelectric devices\cite{Snyder2008, Li_review2022}, thermally insulating coatings\cite{Vassen2010}). Solving the problems of heat flow control is impossible without a detailed understanding of the processes that determine the effective thermal conductivity in heterogeneous nanostructures.
	
	In this paper, we present a consistent microscopic approach based on the analytical solution of the Boltzmann transport equation to describe the processes that contribute to the removal of heat from the heated part of a layered system to its colder part.
	We turn to the microscopic theory since the macroscale Fourier law cannot be used to calculate the effective cross-plane thermal conductivity of layered structures if the layer thickness is less than the length at which phonon thermalization occurs. 
	Inside such layers, phonons are not in local thermodynamic equilibrium, and this makes it impossible to introduce a local temperature.\cite{Chen2021}
	
	Previously, the cross-plane thermal conductivity of layered structures was analyzed based on the Boltzmann equation in terms of the phonon intensity, that is, the phonon energy flux in a specific direction.
	Such a description usually uses the assumption that all phonon modes have the same relaxation time (the gray medium approximation)\cite{Majumdar, Chen, Ordonez}. In several recent papers (see, for example, Ref.\cite{Hua&Minnich} and the literature cited therein), the $\tau$-approximation is also used, although it is assumed that each phonon mode has its own relaxation time. Note that for thin layers and large temperature drops, the use of the collision integral in the $\tau$-approximation with a temperature-dependent equilibrium phonon distribution function is not justified, since it is impossible to enter the local temperature correctly.

	Our microscopic approach to calculating the cross-plane thermal conductivity in a system of metal layers separated by dielectric interlayers is based on the Boltzmann equation for the phonon distribution function and the condition of thermalization of the electronic subsystem in metal layers. The situation when electrons in a metal are thermalized, but phonons are not, is realized at sufficiently low temperatures, at which the characteristic time of electron-electron collisions is much shorter than the time of collisions of electrons with phonons. A simple estimate shows that in pure metals such a picture takes place at temperatures $T< k_B\Theta_{D}^2 /\varepsilon_F \sim $ 1 K. Here $k_B$ is the Boltzmann constant, $\Theta_{D}$ is the Debye temperature, and $\varepsilon_F$ is the Fermi energy. Note that in dirty metals, the enhancement of the electron-electron interaction and the weakening of the electron-phonon interaction can increase the estimate of the upper-temperature limit by an order of magnitude.\cite{Gershenzon}
	
	Interest in the region of helium temperatures is caused, in particular, by the fact that as the temperature decreases, the level of thermal noise in electronic devices decreases. This reduction in noise allowed the development of  highly sensitive hot electron nanobolometers\cite{Wei2008, Karasik2011, Kokkoniemi} and single photon detectors.\cite{Goltsman2001, Eisaman2011} In our opinion, the results obtained below can be useful for designing and analyzing the operating modes of low-temperature electronic devices with a layered structure.

	The paper is organized as follows. In Section~\ref{s2}, we formulate a kinetic approach to cross-plane heat transfer in a system with two metal layers. Section~\ref{s3} presents calculations of the effective transverse thermal conductivity in a layered system for the cases of thick and thin metal layers. The results obtained for a system with two metallic layers are generalized to the cases of a system with an arbitrary number of layers and a superlattice.
	Section~\ref{s4} contains a discussion of the results and Section~\ref{s5} contains conclusions.

	\section{Microscopic approach to the analysis of cross-plane heat transfer in a system with two metal layers}\label{s2}

	\begin{figure}
		\includegraphics [width=6cm ]{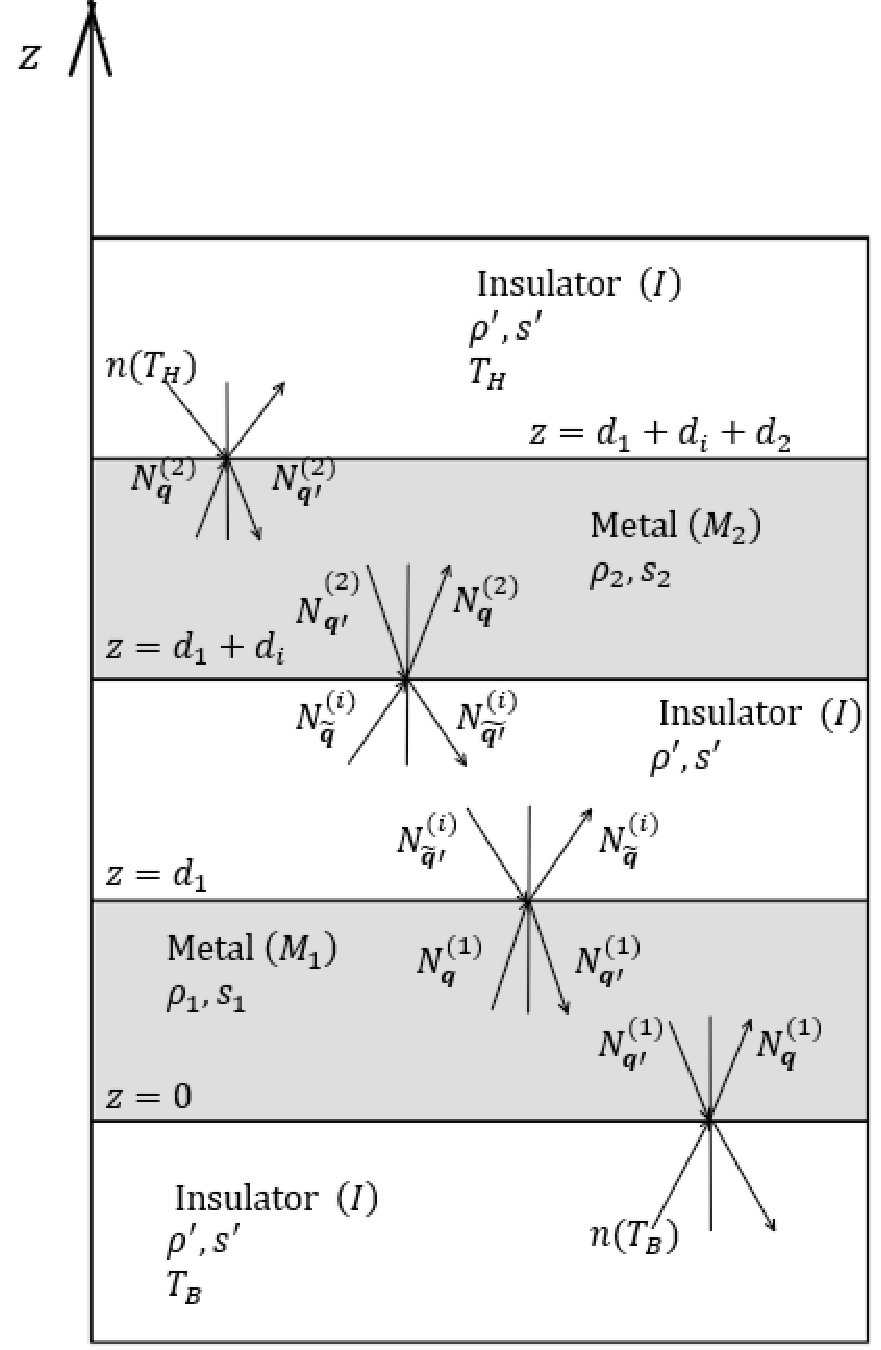}		
		\caption{ 
			Refraction and reflection of phonon modes at  metal/dielectric boundaries in the ${I/M}_1/I/M_2/I$ structure. The directions of wave vectors of the phonons and the number of filling the corresponding phonon states are shown. $T_B$ is the substrate temperature, $T_H$ is the temperature of the upper dielectric plate. $s_{1(2)}$    $(s')$ are velocities of the longitudinal sound in the metals (dielectric), $\rho_{1,2}$ is the densities of the metal layers, $\rho'$ is the density of the dielectric. Metals $M_1$ and $M_2$ are believed to be different, and the dielectric substrate material coincides with the material of layer $I$ and the material of the upper dielectric plate.
		}
		\label{Figure}
	\end{figure}

	An approach to interlayer heat transfer based on the Boltzmann equation for the phonon distribution function makes it possible to find heat flux in layered structures at large temperature drops and layer thicknesses exceeding the characteristic phonon wavelength.
	Consider a layered system containing two metal layers (see Fig.~\ref{Figure})  and assume that phonons have only the longitudinal vibration branch.
	In the steady case, we write the Boltzmann transport equation for the phonon distribution function $N_\mathbf{q}^{\left(k\right)}$   in the metallic layer $M_k$\  \ $\left(k=1,\ 2\right)$  as
	
	\begin{equation}\label{1} 
		s_{kz}\frac{dN_\mathbf{q}^{\left(k\right)}}{dz}=I_{pe}\left(N_\mathbf{q}^{\left(k\right)}\right).
	\end{equation}
	Here $s_{kz}$  is the projection of the phonon velocity on the $z$-axis, perpendicular to the layers, $\mathbf{q}$ is the phonon wave vector, and ${I}_{pe}$ is the phonon-electron collision integral. In the case of the Fermi electron distribution function, the general form of the collision integral ${I}_{pe}$ is significantly simplified and takes the form of the $\tau$-approximation that describes the relaxation of the phonon distribution function in the layer $M_k$ to the Bose distribution with local electron temperature:
	
	\begin{equation}\label{2} 
		I_{pe}\left(N_\mathbf{q}^{\left(k\right)}\right)=\nu_k\left[n\left(T_k\right)-N_\mathbf{q}^{\left(k\right)}\left(z\right)\right].
	\end{equation}
	Here $n(T_k)=\left[\exp{\left(\frac{{\hbar\omega}_q}{k_B T_k}\right)}-1\right]^{-1}$ is the Bose distribution function with $z$-dependent electron temperature $T_k$.
	
	In pure metals, the frequency of phonon-electron collisions is given by the following expression:

	\begin{displaymath}\label{3}
		\nu_k=\frac{m_k\mu_k\omega_q}{2\pi\hbar^3\rho_ks_k}
	\end{displaymath}
	where $m_k$ is the effective electronic mass,  $\mu_k$  is the deformation potential constant, $\rho_{k}$ is the density,  and $s_k$ is the velocity of the longitudinal sound in the layer $M_k$. Note that the phonon frequency $\omega_q$ does not change when a phonon passes from one layer to another. 
	The expression for $\nu_k$ can be generalized to the case of dirty metals\cite{Bergmann, Bezuglyj1997} and the phonon-electron collision frequency can be rewritten as

	\begin{displaymath}\label{3-a}
		\nu_{k}=\frac{m^{2} \mu_{kr}^2}{2\pi\hbar^4 \rho_k s_k}(\hbar \omega_q)^{1+r}.
	\end{displaymath}
	Here  $\mu_{kr}$ is a constant characterizing the interaction of electrons with the lattice. The number $r$ describes the effect of the elastic scattering of electrons by defects in the crystal lattice on the frequency of inelastic collisions of electrons and phonons.  By the experimental data, we will assume that the value of $r$ is an arbitrary number from -1 to 1.\cite {Lin&Bird}

	To formulate the boundary conditions, we turn to Fig.~\ref{Figure}, which illustrates the processes of reflection and refraction of phonon modes at layer boundaries. In the metal layer $M_1$ near the boundary with the substrate, the distribution function of phonons having a positive $z$-component of the wave vector contains two contributions. One of them is determined by phonons coming from the substrate, and the other is due to the phonons of the metal, reflected from the boundary.
	
	\begin{equation}\label{4} 
		N_\mathbf{q}^{\left(1\right)}\left(0\right)=\alpha_1n\left(T_B\right)+
		\beta_1N_{\mathbf{q}^\prime}^{\left(1\right)}\left(0\right).
	\end{equation}
	Here $\alpha_1$, $\beta_1$ are the coefficients of transmission and reflection for the interface between the metals $M_1$ and the dielectric. Everywhere in the boundary conditions, the wave vectors $\mathbf{q}$ and $\mathbf{q}^\prime$ represent phonons that have a positive or negative  $z$-component of the wave vector.
	The interface transmission coefficient for phonons incident on the interface at an angle $\theta$ is determined by the acoustic impedances of the adjacent media $Z = \rho s$ and  $Z' = \rho' s'$:
	
	\begin{displaymath}\label{4a}
		\alpha(\theta)= {\frac{4ZZ^{\prime}{\rm cos}\theta{\rm cos}\theta '}{(Z{\rm cos}\theta ' + Z^{\prime}{\rm cos} \theta)^2}},
	\end{displaymath}
	where $\theta '$ is the angle of refraction.\cite{Little, Swartz&Pohl-1989, Chen2022}

	Conditions on the boundaries  $z=d_1$  and $z=d_1+d_i$ are written similarly to the relation~(\ref {4}). For $z=d_1$ we have
	
	\begin{equation}\label{5} 
		N_{\widetilde{\mathbf{q}}}^{\left(i\right)}=
		\alpha_1 N_\mathbf{q}^{\left(1\right)}\left(d_1\right)+
		\beta_1 N_{\widetilde{\mathbf{q'}}}^{\left(i\right)},\ \ 
	\end{equation}
	
	\begin{equation}\label{6}
		N_{\mathbf{q'}}^{\left(1\right)}\left(d_1\right)=
		\alpha_1 N_{\widetilde{\mathbf{q'}}}^{\left(i\right)}+
		\beta_1 N_\mathbf{q}^{\left(1\right)}\left(d_1\right),
	\end{equation}
	whereas at $z=d_1+d_i$
	
	\begin{equation}\label{7}
		N_\mathbf{q}^{\left(2\right)}\left(d_1+d_i\right)=
		\alpha_2 N_{\widetilde{\mathbf{q}}}^{\left(i\right)}+
		\beta_2 N_{\mathbf{q'}}^{\left(2\right)}\left(d_1+d_i\right),
	\end{equation}

	\begin{equation}\label{8}
		N_{\widetilde{\mathbf{q'}}}^{\left(i\right)}=
		\alpha_2 N_{\mathbf{q'}}^{\left(2\right)}(d_1+d_i)+
		\beta_2 N_{\widetilde{\mathbf{q}}}^{\left(i\right)}.
	\end{equation}
	It should be noted that the boundary conditions are written on the assumption that the metals $M_1$ and $M_{2}$ are different, but the materials of the substrate, top plate, and interlayer are the same. As we neglect the scattering of phonons in the layer $I$, the filling numbers $N_{\widetilde{\mathbf{q}}}^{\left(i\right)}$ are not dependent on the coordinate $z$.
	The last boundary condition is

	\begin{equation}\label{9}
		N_{\mathbf{q'}}^{\left(2\right)}\left(d_1+d_i+d_2\right)=
		\alpha_2 n\left(T_H\right)+
		\beta_2 N_\mathbf{q}^{\left(2\right)}\left(d_1+d_i+d_2\right).	              	
	\end{equation}
	Here $\alpha_2$ and $\beta_2$ are the coefficients of transmission and reflection for the interface between the metal $M_2$ and dielectric.	
	
	The equation 
	
	\begin{equation}\label{10}
		\frac{{dN\ }_\mathbf{q}^{(k)}}{dz}+\frac{1}{l_k}{N\ }_\mathbf{q}^{(k)}=\frac{1}{l_k}n\left(T_k(z)\right)
	\end{equation}   
	
	has the following solution:	
	
	\begin{equation}\label{11}
		\begin{split}
			{N\ }_\mathbf{q}^{\left(k\right)}\left(z\right)={C\ }_\mathbf{q}^{\left(k\right)}e^{-z/l_k}+ 
			\\ +\frac{1}{l_k}\int_{z_{ik}}^{z}{e^{-\left(z-z^\prime\right)/l_k}n\left(T_k\left(z^\prime\right)\right)dz^\prime,}
		\end{split}
	\end{equation}
	where $z_{i1}=0$, $z_{i2}=d_1+d_i$, and $l_k(=|s_{kz}|/\nu_k)$  is the free path length of the phonon in the layer $M_k$.
	
	Note that the equation for ${N}_{\mathbf{q'}}^{(k)}\left(z\right)$ differs from Eq.~(\ref {10}) by the sign before the first term on the left side.
	The solution of the equation for ${N}_{\mathbf{q'}}^{(k)}\left(z\right)$ has the following form:
	
	\begin{equation}\label{12}
		\begin{split}{N}_{\mathbf{q'}}^{(k)}\left(z\right)={C}_{\mathbf{q'}}^{(k)}e^{z/l_k}+ \\
			+\frac{1}{l_k} \int_{z}^{z_{fk}}e^{(z-z^\prime)/l_k}n\left(T_k\left(z^\prime\right)\right)dz^\prime,	 
		\end{split}
	\end{equation}
	where $z_{f1}=d_1$, $z_{f2}=d_1+d_i+d_{2}.$	
	
	Constants ${C}_\mathbf{q}^{(k)}$ and ${C}_{\mathbf{q'}}^{(k)}$ are determined by the boundary conditions~(\ref {4}-\ref {9}). For these constants, the following rather cumbersome expressions are obtained:
	
	\begin{widetext}
		\begin{subequations}
			\label{12} 
			
			\begin{eqnarray}
				\begin{aligned}
					&
					{C}_\mathbf{q}^{(1)}=\frac{e^{d_1/l_1}}{D}\{e^{d_1/l_1}(\alpha_1 n(T_B)+\beta_1J_{11})
					(e^{2d_2/l_2}-E_2\beta_2)+\beta_1J_{12}\beta_2(E_{1}^{2}-E_{2}^{2})+
					{} \\
					& {}
					+\beta_1 E_1 e^{d_2/l_2}[(\alpha_{2} n(T_H)+\beta_{2} J_{22})+ e^{d_2/l_2}(E_2 J_{12}+E_1 J_{21})]\},          
				\end{aligned}    \label{12a}
				\\
				\begin{aligned}
					&
					{C}_{\mathbf{q'}}^{(1)}=\frac{1}{D}\{\beta_2(E_{1}^{2}-E_{2}^{2})(J_{12}e^{d_1/l_1}+(\alpha_{1}n(T_B)+\beta_{1}J_{11}))+E_1 e^{d_2/l_2} e^{d_1/l_1} \times
					{} \\
					& {}
					(\alpha_{2}n(T_H)+\beta_{2}J_{22})+e^{2d_2/l_2}[e^{d_1/l_1}(E_2 J_{12}+E_1 J_{21})+E_2(\alpha_{1} n(T_B)+\beta_{1} J_{11})]\},
				\end{aligned} \label{12b}
				\\
				\begin{aligned}
					&
					{C}_\mathbf{q}^{(2)}e^{-(d_1+d_i)/l_2}=\frac{e^{d_2/l_2}}{D}\{E_1e^{d_1/l_1}e^{d_2/l_2}(\alpha_1 n(T_B)+\beta_1J_{11})+\beta_1(E_{1}^2-E_{2}^2)\times 
					{} \\
					& {}	   (J_{21}e^{d_2/l_2}+\alpha_2n\left(T_H\right)+\beta_2J_{22})+e^{2d_1/l_1}[e^{d_2/l_2}(E_1J_{12}+E_2J_{21})+E_2(\alpha_2n\left(T_H\right)+\beta_2J_{22})]\},
				\end{aligned}\label{12c}
				\\
				\begin{aligned}
					&
					{C}_{\mathbf{q'}}^{(2)}e^{(d_1+d_i)/l_2}={\frac{1}{D}}\{(\alpha_2 n (T_H)+\beta_2 J_{22})e^{d_2/l_2}(e^{2d_1/l_1}-E_2\beta_1)+
					{} \\
					& {}	  
					+e^{2d_1/l_1}\beta_2(E_1J_{12}+E_2J_{21})+{e^{d_1/l_1}}\beta_2 E_1(\alpha_1 n(T_B)+\beta_1 J_{11})+\beta_1 \beta_2 J_{21}(E_{1}^2-E_{2}^2)\},
				\end{aligned}\label{12d}
			\end{eqnarray}
		\end{subequations}	
	\end{widetext}	
	where
	\begin{displaymath}
		E_1=\frac{{\alpha_1\alpha}_2}{(1-\beta_1\beta_2)},\ \ E_2=\frac{\alpha_1\beta_2+\beta_1\alpha_2}{(1-\beta_1\beta_2)}.
	\end{displaymath}  
	The determinant of the system of linear equations for the coefficients ${C}_\mathbf{q}^{(k)}$ and ${C}_{\mathbf{q'}}^{(k)}$ is
	
	\begin{equation*}
		\begin{split}
			D=e^{{2d}_1/l_1}e^{{2d}_2/l_2}-\beta_1 E_2e^{{2d}_2/l_2}-\beta_2 E_2 e^{{2d}_1/l_1}- \\- \beta_1\beta_2(E_{1}^2-E_{2}^2).
			\end {split}
		\end{equation*}
		Eqs.~(\ref {12}) also contain the constants $J_{ik}$, which are defined by the following equalities:
		
		\begin{subequations}
			\label{13} 
			\begin{eqnarray}
				J_{11}= \frac{1}{l_1} \int_{0}^{d_1}e^{-z^\prime/l_1}n\left(T_1\left(z^\prime\right)\right)dz^\prime,\label{13a}
				\\
				J_{12}=\frac{1}{l_1} \int_{0}^{d_1} e^{{-(d_1-z}^\prime)/l_1} n(T_1(z^\prime))dz^\prime,\label{13b}
				\\
				J_{21}=\frac{1}{l_2} \int_{z_{i2}}^{z_{f2}} e^{(z_{i2}-z^\prime)/l_2} n(T_2(z^\prime))dz^\prime,\label{13c}
				\\
				J_{22}=\frac{1}{l_2} \int_{z_{i2}}^{z_{f2}}e^{-(z_{f2}-z^\prime)/l_2} n\left(T_2\left(z^\prime\right)\right)dz^\prime.\label{13d}
			\end{eqnarray}
		\end{subequations}

		\section{Effective cross-plane thermal conductivity}\label{s3}
		
		Under stationary conditions, the transverse heat flux in a layered system is a conserved quantity, and therefore it can be calculated for any coordinate $z$.
		Provided that the electrons are in local thermodynamic equilibrium, the $z$-projection of the heat flux in the layer $M_1$ is represented by the expression
		
		\begin{equation}\label{14}
			Q_z=-k_{e1}[T_{1}(z)]\frac{dT_{1}}{dz}+\int_{q_{z}>0}\frac{d^3 
				q}{(2\pi)^3}\hbar\omega_{q}s_{1z}
			\biggl[N_{\bf q}^{(1)}-N_{\bf q'}^{(1)}\biggl],
		\end{equation}
		where $k_{e1}$ is the electronic thermal conductivity in the layer $M_1$.\cite{Abrikosov} The second term on the right-hand side is the fraction of the heat flux carried by phonons.
		The substitution of the phonon distribution function found in Section~\ref{s2} into Eq.(\ref {14}) leads to a cumbersome expression, which is hardly suitable for elucidating the physical picture of heat transfer in layered structures. To obtain such a picture, it is necessary to pass from the general case to the limiting cases of thin and thick metal layers.

		\subsection{Thin metal layers}\label{ss3.1}

		Let us turn to the case of thin $M$-layers, that is, we will assume that $d_1\ll l_1$, and $d_2\ll l_2$.
		In the zeroth approximation in $d_1/l_1$, the phonon distribution function in layer $M_1$ is given by the following equalities:

		\begin{equation}\label{15}
			{N}_\mathbf{q}^{(1)}=\frac{1}{D}\{\alpha_1n (T_B)-E_2\alpha_1 \beta_2 n(T_B)+E_1\beta_1\alpha_2n (T_H)\},
		\end{equation} 
		
		\begin{eqnarray}\label{16}
			{{N}_{\mathbf{q}\prime}}^{(\mathbf{1})}=\frac{1}{D}\{E_1\alpha_2n (T_H)+E_2\alpha_1n (T_B)+\nonumber \\+\alpha_1\beta_2 n(T_B)(E_{1}^2-E_{2}^2)\},
		\end{eqnarray}
		where
		\begin{displaymath}
			D=1- \beta_1 E_2-\beta_2 E_2-\beta_1\beta_2 (E_{1}^2-E_{2}^2).
		\end{displaymath}

		Neglecting the electronic contribution to the thermal conductivity of the metal layer (since phonons in thin $M$-layers do not have time to transfer energy to electrons), we have

		\begin{equation}\label{17}
			Q_z =\int_{q_z>0}\frac{d^3q}{{(2\pi)}^3}\hbar\omega_qs_{1z}[N_\mathbf{q}^{(1)}-N_{\mathbf{q}\prime}^{(1)}].   
		\end{equation}       
		Substituting ${N}_\mathbf{q}^{\left(1\right)}$ and ${N}_{\mathbf{q}\prime}^{(1)}$ into Eq.(\ref {17}) gives the following expression for the heat flux for an arbitrary temperature difference between the heated top layer and the cold substrate: 
		
		\begin{equation}\label{18}
			Q_z = -\frac{\pi^2}{120}\frac{k_B^4}{\hbar^3s_1^2}(T_{H}^4-T_{B}^4)\left\langle\frac{E_1}{1+E_2}\right\rangle.
		\end{equation}
		Here the angle brackets denote the averaging over the angles of incidence $\theta_1$
		
		\begin{displaymath}
			\left\langle f\right\rangle = \int_{0}^{\pi/2}{\rm sin (2\theta_1) f(\theta_1)}d \theta_1.
		\end{displaymath}

		The effective thermal conductivity of the layered structure with two metal layers $M_1$ and $M_2$ is determined by the relation
		
		\begin{equation}\label{19}
			\kappa_{eff}=\frac{\left|Q_z\right|(d_1+d_i+d_2)}{T_H-T_B}
		\end{equation}   
		provided $T_H-T_B<<T_B$. For a system with two thin metal layers, we have
		
		\begin{equation}\label{20}
			\kappa_{eff}=\frac{\pi^2}{30}\frac{k_B^4 T_{B}^3}{\hbar^3s_1^2}\left\langle\frac{E_1}{1+E_2}\right\rangle (d_1+d_i+d_2).
		\end{equation}   
		
		Let us clarify the physical meaning of the result (\ref {20}) and generalize it to systems with $n$ thin metal layers.
		We denote the probability of the phonon transmission through a two-layer system with  thin $M$-layers by $T^{(2)}$. In the absence of absorption of phonons in the $M$-layers,
		
		\begin{widetext}
			\begin{equation}\label{21} T^{\left(2\right)}=T_1^{\left(1\right)}T_2^{\left(1\right)}+T_1^{\left(1\right)}T_2^{\left(1\right)}R_1^{\left(1\right)}R_2^{\left(1\right)}+T_1^{\left(1\right)}T_2^{\left(1\right)}\left(R_1^{\left(1\right)}R_2^{\left(1\right)}\right)^2+\ldots=\frac{T_1^{\left(1\right)}T_2^{\left(1\right)}}{1-R_1^{\left(1\right)}R_2^{\left(1\right)}},
			\end{equation}   
		\end{widetext}
		where $T_k^{\left(1\right)}$ is the probability of the phonon transmission through one metal layer $M_k$ and $R_k^{\left(1\right)}$ is  the probability of phonon reflection from a single metal layer $M_k$. Similarly to the previous formula for the probability $T^{\left(2\right)}$, we can write
		
		\begin{equation}\label{22}
			T_k^{\left(1\right)}=\alpha_k^2+\alpha_k^2\beta_k^2+\alpha_k^2\beta_k^4+\ldots=\frac{\alpha_k}{1+\beta_k}\ 
		\end{equation} 
		
		The probability of phonon reflection from a single metal layer $M_k$ is
		
		\begin{equation}\label{23}
			R_k^{\left(1\right)}=\frac{{2\beta}_k}{1+\beta_k}.
		\end{equation}
		
		From this, we have that the probability of the phonon transmission through a two-layer system with  thin $M$-layers is given by
		
		\begin{equation}\label{24}
			T^{(2)}=\frac{E_1}{1+E_2}.
		\end{equation}
		This expression is the same as the expression in angle brackets in Eq.(\ref {18}). Thus, Eq.(\ref {18}) includes the probability of the phonon transmission through a two-layer system $T^{(2)}$. This means that $\kappa_{eff}$ of a system with $n$ metal layers must be determined by a similar transmission probability $T^{(n)}$.
		
		Since in a multilayer system, usually, all metal layers are made of the same metal, and dielectric (or semiconductor) layers are made of the same dielectric (semiconductor), the probability $\alpha_1{=\alpha}_2=\alpha$ and  $\beta_1=\beta_2=\beta$.
		In this case,
		\begin{equation}\label{25}
			T^{\left(1\right)}=\frac{\alpha}{1+\beta},\ \ \ T^{\left(2\right)}=\frac{\alpha}{1+3\beta}.
		\end{equation}
		It can be shown that for an $n$–layered system, the transmission probability is
		
		\begin{equation}\label{26}
			T^{\left(n\right)}=\frac{\alpha}{1+(2n-1)\beta}.
		\end{equation}
		
		Hence, the expression for the heat flow in the $n$-layer system has the form 
		
		\begin{equation}\label{27}
			Q_z = -\frac{\pi^2}{120}\frac{k_B^4}{\hbar^3\ s^2}\left(T_H^4-T_B^4\right)\left\langle\frac{\alpha}{1+(2n-1)\beta}\right\rangle.   
		\end{equation}

		The effective thermal conductivity of a multilayer system with $n$ identical thin $M$-layers and $n-1$ identical dielectric interlayers is

		\begin{equation}\label{28}
			\kappa_{eff}=\frac{\pi^2}{30}\frac{k_B^4T_B^3}{\hbar^3\ s^2}\left\langle\frac{\alpha}{1+(2n-1)\beta}\right\rangle \cdot \left[  nd_m+(n-1)d_i\right], 
		\end{equation}
		where	$d_m$ is the thickness of the $M$-layer, $s$ is the velocity of longitudinal sound in the $M$-layer, and $d_i$ is the thickness of the $I$-layer.
		If $n \gg\ 1$, 
		
		\begin{equation}\label{29}
			\kappa_{eff}=\frac{\pi^2}{60}\frac{k_B^4T_B^3}{\hbar^3\ s^2}\left\langle\frac{\alpha}{\beta}\right\rangle(d_m+d_i).   
		\end{equation}
		Thus, the effective thermal conductivity of the multilayer system with thin metallic layers is determined by the period of the system and does not depend on the number of layers. If we compare Eq.(\ref {29}) with the well-known expression for phonon thermal conductivity $\kappa_p \sim c_{p} s l_p$, where $c_p$ is the phonon heat capacity, and $l_p$ is the mean free path of phonons, we get $l_{eff} \sim \left\langle \frac{\alpha}{\beta}\right\rangle (d_m+d_i)$;  $l_{eff}$  is the effective mean free path of phonons.

		\subsection{Thick metal layers}\label{ss3.2}
		
		Above, we obtained an expression for the heat flux in the case of thin $M$-layers. 
		Let us now find an expression for the heat flux in the case of thick $M$-layers.
		To do this, it is necessary to use expressions Eq.~(\ref {11}) and Eq.~(\ref {12}) with the coefficients $C_\mathbf{q}$ calculated in the limit $d_1\gg l_1$, $d_2 \gg l_2$.
		For the layer $M_1$, we have

		\begin{widetext}
			\begin{subequations}
				\label{30} 
				\begin{eqnarray}
					{N}_\mathbf{q}^{(1)}(z)=e^{-z/l_1}\cdot[\alpha_1 n(T_B)+\beta_1 J_{11}]+\frac{1}{l_1}\int_{0}^{z}e^{-(z-z^\prime)/l_1}n(T_1(z^\prime))dz^\prime,	\label{30a}
					\\
					{N}_{\mathbf{q}^\prime}^{(1)}(z)=e^{(z-d_1)/l_1}\cdot(E_2 J_{12}+E_1 J_{21})+\frac{1}{l_1}\int_{z}^{d_1}e^{(z-z^\prime)/l_1}n(T_1(z^\prime))dz^\prime.\label{30b}
				\end{eqnarray}
			\end{subequations}
		\end{widetext}
		With a small temperature drop in the layer $M_1$, the function $n(T_1(z'))$ can be expanded near the point $z$: 
		
		\begin{displaymath}
			n(T_1(z^\prime)) \approx n(T_1(z)-\frac{dn}{dT_1}\frac{dT_1(z)}{dz}(z-z^\prime).
		\end{displaymath}   
		In parts of the metal layer $M_1$ that are not too close to the edges of the layer, that is, for $z\gg l_1$ and $d_1-z \gg l_1$, the exponents $e ^{-z/l_1} $ and $e^{-(d_1-z)/l_1}$ are small.
		As a consequence, the components in ${N\ }_\mathbf{q}^{\left(1\right)}\left(z\right)$ and in ${N\ }_{\mathbf{q}^\prime}^{\left(1\right)}\left(z\right)$ containing $e^{-z/l_1}$ and $e^{-\left(d_1-z\right)/l_1}$ can be omitted. For the phonon distribution function, we obtain the following expression:

		\begin{equation}\label{31}
			N_{\mathbf{q}({\mathbf{q'})}}^{\left(1\right)}\left(z\right)=n(T_1\left(z\right)) \mp l_1\frac{dn}{d T_1}\frac{dT_1\left(z\right)}{dz}.
		\end{equation}       
		Therefore, the heat flux carried by phonons is
		
		\begin{equation} \label{32}
			Q_{z}^{\left(ph\right)}(z)=
			-2\int_{q_z>0}\frac{d^3q}{{(2\pi)}^3}\hbar\omega_q s_{1z} l_1\frac{dn}{dT_1}\frac{dT_1(z)}{dz}. 
		\end{equation}     
		
		The total heat flux in the $M_1$ layer consists of the electronic and phonon contributions. 
		The z-projection of the total heat flux is
		
		\begin{equation}\label{33}
			Q_{z} = -\kappa_{e1}(T_B)\frac{dT_1}{dz}-\kappa_{p1}(T_B)\frac{dT_1}{dz}.
		\end{equation}     
		where $\kappa_{p1}(T_B)$ is the phonon thermal conductivity of metal $M_1$ at the temperature $T_B$. From the above results, it follows that
		
		\begin{eqnarray}\label{34}
			\kappa_{p1}(T)=2\int_{q_z>0}\frac{d^3q}{\left(2\pi\right)^3}\hbar\omega_q\frac{s_{1z}^2}{\nu_1}\frac{dn_q}{dT}  \nonumber \\ =
			\Gamma (3-r)\zeta (3-r)\frac{(3-r)}{3\pi} \frac{\hbar \rho_1}{m_{1}^{2} \mu_{1r}^{2}}
			k_{B}^{3-r} T^{2-r},
		\end{eqnarray}  
		where $\Gamma$ and $\zeta$ are the gamma function and the Riemann zeta function, respectively. The expression~(\ref {34}) generalizes the well-known relation for the phonon thermal conductivity of pure metals, $\kappa_p \propto T^2$, to the case of dirty metals.

		In the case of thick $M$-layers, the difference $T_H-T_B$ can be represented as the sum of the temperature drops at the $z= 0$ and $z= d_1 + d_i + d_2$ interfaces, as well as on the $M_1$ and $M_2$ layers and the $I$ interlayer.
		The temperature drops at the boundaries $z=d_1$ and $z=d_1+d_i+d_2$ are determined by the thermal resistances of the boundaries as $\left|Q\right|R_{th,k}$, where $k=1,2$.
		The thermal resistance of the boundary is derived from Little's results\cite{Little}. In our notations        
		
		\begin{equation}\label{35}
			R_{th,k}=\left[\frac{\pi}{30}\frac{k_B^4T_B^3}{\pi^3s_k^2}\cdot <\alpha_k>\right]^{-1}.
		\end{equation}

		The temperature drop on the $M_k$ layer is determined by the heat flux as $\left|Q_z\right|d_k/\kappa_{T,k}$,
		where $\kappa_{T,k}=\kappa_{e,k}+\kappa_{p,k}$.

		The temperature difference $T_2\left(d_1+d_i\right)-T_1\left(d_1\right)$ can be found from the expression for the heat flow in the dielectric interlayer $I$.   Calculations lead to the result $\left|Q_z\right|=R_I [{T}_2\left(d_1+d_i\right)-T_1\left(d_1\right)]$, where the thermal resistance of the dielectric interlayer is
		
		\begin{equation}\label{36}
			R_{thI}=\left[\frac{\pi^2}{30}\frac{k_B^4T_B^3}{\hbar^3s_i^2}\cdot \left \langle \frac{\alpha_1\alpha_2}{1-\beta_1\beta_2}\right \rangle  \right ]^{-1}.
		\end{equation}   
		Here ${s}_i$ is the speed of longitudinal waves in the interlayer $I$.
		Thus, we have the following expression for the effective thermal conductivity in a layered system with two metal layers:
		
		\begin{equation}\label{37}
			\kappa_{eff}=\frac{d_1+d_i+d_2}{R_{th1}+\ R_{th2}+R_{thI}+\frac{d_1}{\kappa_{T,1}}+\frac{d_2}{\kappa_{T,2}}\ }.
		\end{equation}
		This expression is easily generalized to the case of $n$  metal layers. We obtain  
		
		\begin{equation}\label{38}
			\kappa_{eff}=\frac{\sum_{l=1}^{n}d_l+\sum_{l=1}^{n-1}d_{i,l}}{R_{th1}+\ R_{th2}+\sum_{l=1}^{n-1}R_{thI,l}+\sum_{l=1}^{n}\frac{d_l}{\kappa_{T,l}} }.
		\end{equation}      
		Here $d_{i,l}$ is the thickness of the dielectric interlayer with the number $l$, and $R_{thI,l}$ is its thermal resistance.

		In the case of a multilayer system ($n \gg 1$) with identical metal layers of thickness $d_m$ and identical dielectric interlayers of thickness $d_i$, the expression for the effective cross-plane thermal conductivity is reduced to

		\begin{equation}\label{39}
			\kappa_{eff}=\frac{d_m+d_{i}}{R_{thI}+\frac{d_m}{\kappa_{T,m}}}.
		\end{equation}      
		Here $\kappa_{T,m}$ is the total thermal conductivity of the metal layer, and in expression~(\ref {36}) for $R_{thI}$ it is necessary to put $\alpha_1=\alpha_2=\alpha $ and $\beta_1=\beta_2=\beta$. This result differs from the well-known expression for the cross-plane thermal conductivity of a superlattice, obtained on the basis of the concepts of series-connected thermal resistances of metallic and dielectric layers, as well as the boundaries between them.\cite{Capinski, Aksamija, Rawat} The difference arises because we do not assume that the phonons in the dielectric layers are thermalized. This situation is quite real for the region of low temperatures we are considering.

		\section{Discussion }\label{s4}
		
		Heat transfer in layered systems "metal-dielectric" or "metal-semiconductor" is considered microscopically based on the analytical solution of the Boltzmann equation for the phonon distribution function, provided that electrons are thermalized in metal layers. As noted in Section~\ref{s1}, this condition is satisfied at sufficiently low temperatures, when the frequency of phonon-electron collisions is much lower than the frequency of electron-electron collisions. Another feature of the model is that only one branch of the phonon spectrum is taken into account. It is in this assumption that the boundary conditions are written in Section~\ref{s2}. Although, within the framework of the presented microscopic approach, all vibration branches can be taken into account, such a generalization would lead to extremely cumbersome and poorly transparent expressions. Note that in the approximation used, we obtained the final results in terms of thermal conductivities of metal layers and thermal resistances of interfaces and dielectric layers, that is, in a form that does not depend on the number of branches of the phonon spectrum taken into account.
		This form of results makes it possible to compare theoretical results with experiments.

		When analyzing heat transfer in layered structures, it is important to highlight the size effect, which has common features with the size effect when heat is removed from a metal film to a dielectric substrate.\cite{ShklovskiiJETP1980}  If the thickness of the metal layer is less than or of the order of the mean free path of thermal phonons, $d_m \lesssim l_p$, the so-called ballistic heat transfer regime is realized. According to Eq.~(\ref {29}), in this mode, the effective thermal conductivity of the multilayer system is proportional to its period and strongly depends on temperature ($\kappa_{eff}\propto T_{B}^3$). 
		
		In the opposite case ($d_m \gg l_p$), heat transfer has a diffusive character.
		If in the equation~(\ref {39}) $R_{thI} \ll d_m/\kappa_m$ and $d_m\approx d_i$, the effective thermal conductivity $\kappa_{eff}$ is determined by the properties of the metal layers and weakly depends on the period thickness of the layered structure.
		When the temperature is low enough, so that $R_{thI}\gg d_m/\kappa_m$, the effective thermal conductivity is determined by the thermal resistance of the dielectric layer, and we can say that it is in this case that the layered nature of the multilayer system manifests itself.
		
		The dependence of $\kappa_{eff}$ on the thickness of the superlattice period, similar to the dependence described by the equations (\ref{29}) and (\ref{39}), was experimentally observed at room temperature (see Fig.~4 in Ref.~\cite{Saha2016}). Note that the region of transition from the ballistic regime to the diffusion regime can be used to estimate the mean free path of thermal phonons in the metal layer. Based on the results given in Ref.~\cite{Saha2016}, it can be obtained that for a metal layer of (Ti,W)N at room temperature, $l_p \approx$ 30 nm.
		
		Finally, we note that the microscopic approach developed in Section~\ref{s2} and Section~\ref{s3} can be applied to the calculation of the effective thermal conductivity of layered systems consisting of layers of a dielectric and a ferromagnetic insulator.\cite{Shklovskij&Bezuglyj2021} In this case, the role of electrons passes to magnons. Since the magnon-phonon interaction is much weaker than the electron-phonon interaction, the temperature range of applicability of the developed formalism will be much wider than for the "metal-dielectric" layered systems considered here.

		\section{Conclusions}\label{s5}
		
		The problem of effective cross-plane thermal conductivity of structures containing two metal layers is analyzed in detail based on solving the Boltzmann transport equation for the phonon distribution function.   The results obtained are generalized to multilayer "metal-dielectric" structures and the size effect in transverse heat transfer in such structures is emphasized. If the thickness of the metal layers in the superlattice is much less than the phonon mean free path, the effective cross-plane thermal conductivity is proportional to the period thickness of the superlattice. In the opposite case,  such thermal conductivity weakly depends on the period thickness of the superlattice.
		
		The effect of phonon reflection from the metal-dielectric interfaces on the cross-plane thermal conductivity increases with decreasing temperature. Thus, it is at low temperatures that the layered nature of the system can lead to a significant decrease in the cross-plane thermal conductivity.
		
		\section*{Acknowledgment}
		We thank Professor R.V. Vovk for helpful discussions and support of our work.

		{}

	\end{document}